\documentclass{PoS}

\def\beq {\begin{equation}}
\def\eeq {\end{equation}}
\def\bea {\begin{eqnarray}}
\def\eea {\end{eqnarray}}

\def\nn {\nonumber}
\def\vs {\vspace}

\def\lp {\left( }
\def\rp {\right) }
\def\lb {\left[ }
\def\rb {\right] }
\def\lc {\left\{ }
\def\rc {\right\} }
\def\ra {\;\rangle }
\def\la {\langle\; }

\def\rar {\rightarrow}

\def\db {\bar{d}}
\def\ub {\bar{u}}

\def\Ft {\tilde{F}}
\def\Fc {\check{F}}

\def\Lat {\tilde{\Lambda}}

\def\sm {\!-\!}
\def\cd {\!\cdot\!}

\def\uma {(1\!-\!a)}

\def\cL {{\cal{L}}}

\def\cO {{\cal{O}}}

\def\d {\delta}
\def\D {\Delta}

\def\g {\gamma}

\def\m {\mu}

\def\p {\pi}

\def\s {\sigma}

\def\T {\Theta}

\def\bp {\mbox{\boldmath $p$}}
\def\bq {\mbox{\boldmath $q$}}
\def\br {\mbox{\boldmath $r$}}

\def\btau {\mbox{\boldmath $\tau$}}
\def\bnb {\mbox{\boldmath $\nabla$}}
\def\btheta {\mbox{\boldmath $\theta$}}

\title{nuclear interactions\\ and the space-like
structure of the pion}

\ShortTitle{nuclear interactions and the space-like
structure of the pion}

\author{\speaker{M. R. Robilotta}\\
\\
Instituto de F\'{\i}sica, Universidade de S\~{a}o Paulo,\\
C.P. 66318, 05315-970, S\~{a}o Paulo, SP, Brazil \\
      E-mail: \email{robilotta@if.usp.br}}

\abstract{Three instances are discussed in which results produced 
by chiral perturbation theory can be reliably pushed to high space-like 
values of transferred momenta.
\\
{\bf 1. nuclear interactions: \hspace*{1mm}}
At present, expansions are available for about 20 components
of both two- and three-nucleon forces, and the vast majority of them
follows the patterns predicted by chiral symmetry. 
The outstanding exception is $V_C^+$, the isospin independent  
central potential.
Standard calculations show that this $\cO(q^3)$ contribution is about 
10 times larger than the leading $\cO(q^2)$ isospin dependent term $V_C^-$.
In spite of defying counting rules, these results are quite well
supported by phenomenology up to distances smaller than 
1 fm $(\rar |t| \sim 20\; M_\p^2)$.
\\
{\bf 2. nucleon sigma-term: \hspace*{1mm}}
The configuration space nucleon scalar form factor $\Ft_s(r)$
is an important substructure of $V_C^+$, and its integration 
over the entire volume yields $\s_N$, the nucleon $\s$-term.
Perturbative results based on diagrams involving $N$ and $\D$ 
intermediate states vanish at large distances, 
and increase monotonically as one approaches the nucleon center, 
where they can become arbitrarily large.
Assuming that the pion cloud of the nucleon is constructed at the 
expenses of the surrounding condensate, an upper limit for $\Ft_S(r)$ 
can be set at a critical radius 
$R\simeq 0.6$ fm $(\rar |t| \sim 40\; M_\p^2)$, 
where a phase transition takes place.
This mechanism excludes the problematic region and yields 
43 MeV$<\s_N<\;$49 MeV, in agreement with the empirical value 
45$\pm 8$ MeV.
\\
{\bf 3. space-like structure of the pion: \hspace*{1mm}}
The extension of the model for $\s_N$ to the pion describes it
as a Goldstone boson at large distances, surrounded
by a quark-antiquark condensate.
As one moves towards its center, the condensate is gradually destroyed
and a phase transition occurs at a distance 
$R\simeq 0.6$ fm $(\rar |t| \sim 40\; M_\p^2)$. 
When only pion loops are considered, the model depends 
on just $M_\p$ and $F_\p$, and yields 
$\la r^2 \ra_S^\p=0.50$ fm$^2$ and $\bar{l}_4=3.9$.
The inclusion of a scalar resonance of mass $980\,$MeV,
with two known coupling constants, improves these values to 
$\la r^2 \ra_S^\p=0.59$ fm$^2$ and $\bar{l}_4=4.3$,
well within the error bars of the precise estimates
$\la r^2 \ra_S^\p = 0.61 \pm 0.04$ fm$^2$ and  $\bar{l}_4 = 4.4 \pm 0.2$,
produced in 2001 by Colangelo, Gasser and Leutwyler.
In both cases, results are given in terms of
simple analytic expressions.}

\FullConference{6th International Workshop on Chiral Dynamics, CD09\\
		July 6-10, 2009\\
		Bern, Switzerland}

\begin{document}

\section{NUCLEAR INTERACTIONS}

In the last twenty years, our understanding of nuclear interactions
has improved considerably\cite{Epelbaum}, owing to the systematic 
use of chiral perturbation theory (ChPT)\cite{W-ChPT}.
As the masses of the quarks $u$ and $d$ are small, they are treated 
as perturbations in a $SU(2)\times SU(2)$ chiral symmetric lagrangian.
Hadronic amplitudes are then
expanded in terms of a typical scale $q$, set by either pion 
four-momenta or nucleon three-momenta, such that $q\ll 1$ GeV.
This procedure is rigorous and results are written 
as power series in the scale $q$, giving rise to the notion of 
chiral hierearchies.
In most cases, leading order terms come from tree diagrams
and corrections require the inclusion of pion loops.

In spite of all the progress achieved, 
there are still some puzzles in our 
picture of nuclear interactions. 
At present, chiral symmetry has been applied to about 20 components 
of nuclear forces\cite{R-Osaka}, and the {\em predicted} structure for the
most important terms is shown in table \ref{T1}, where $OPE$and $TPE$
stand for {\em one-pion exchange} and {\em two-pion exhange},
$V_i^+$ and $V_i^-$ represent two-body operators 
proportional to either the identity or $\btau^{(1)}\cd \btau^{(2)}$ 
in isospin space, 
$i \rar$ [central$(C)$, spin-orbit$(LS)$, spin-spin$(SS)$, tensor$(T)$],
whereas the notation of ref.\cite{CDR} is used for three-body forces.
Actual results for central components $V_C^\pm$ defy these predictions.

\begin{table}[h]
\begin{center}
\begin{tabular} {|c|ccc|}
\hline
leading	  	& TWO-BODY & TWO-BODY	& THREE-BODY 	   \\ 
contribution				& $OPE$   & $TPE$		& $TPE$		\\ \hline 
$\cO(q^0)$		& $V_T^-, V_{SS}^-$	&&    	  	   \\[1mm] \hline
$\cO(q^2)$		&  & $V_C^-; V_T^+, V_{SS}^+$ &	   \\[1mm]\hline
$\cO(q^3)$		&& $V_{LS}^-, V_T^-, V_{SS}^-; V_C^+, V_{LS}^+$ 
& $W_S,W_P,W_P'$ 	\\[1mm] \hline
\end{tabular}
\label{T1}
\end{center}
\end{table}

\begin{figure}[h]
\begin{center}
\includegraphics[width=0.65\columnwidth,angle=0]{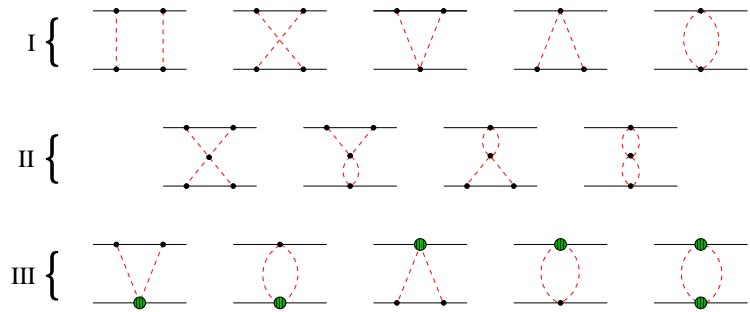}
\caption{Dynamical structure of the two-pion exchange potential.} 
\label{FB1}
\end{center}      
\end{figure}

The chiral two-pion exchange $NN$ potential was studied by our 
group\cite{HR,HRR}, by means of the three families of diagrams displayed in 
fig.\ref{FB1}.
Family $I$ implements the minimal realization of chiral symmetry\cite{RR94} 
and begins at $\cO(q^2)$, 
whereas family $I\!I$ depends on $\p\p$ correlations and 
contributes at $\cO(q^4)$.
Vertices in these familes involve only $g_A$ and $F_\pi$,
and all dependence on other LECs is concentrated in family $I\!I\!I$,
which begins at $\cO(q^3)$.
These LECs can be extracted from subthreshold $\p N$ amplitudes\cite{BL}
and one finds that the components $V_C^-$ and $V_C^+$ are very strongly 
dominated by families $I$ and $I\!I\!I$, respectively.

\begin{figure}[h]
\begin{center}
\hspace*{-4mm}
\includegraphics[width=0.50\columnwidth,angle=0]{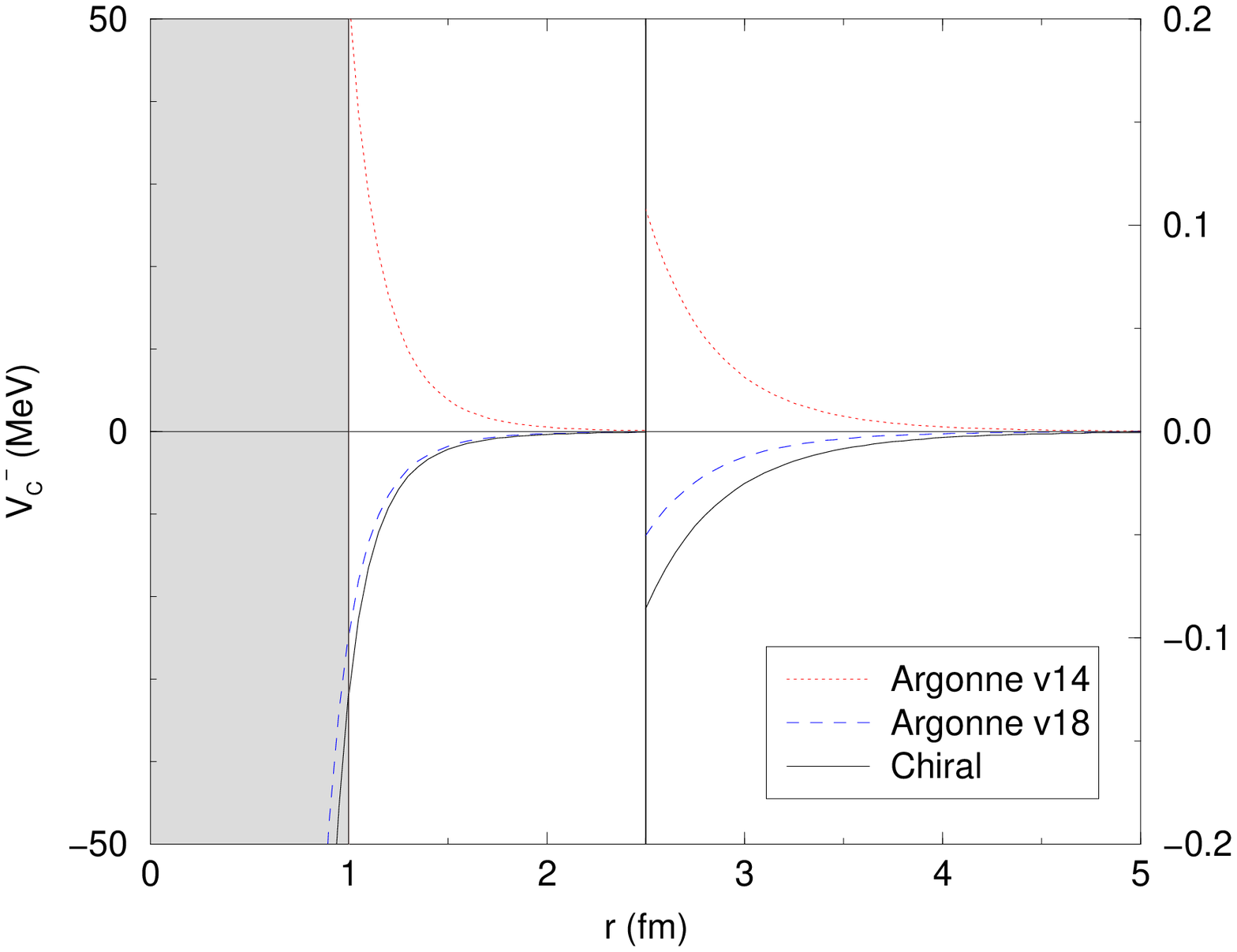}
\includegraphics[width=0.50\columnwidth,angle=0]{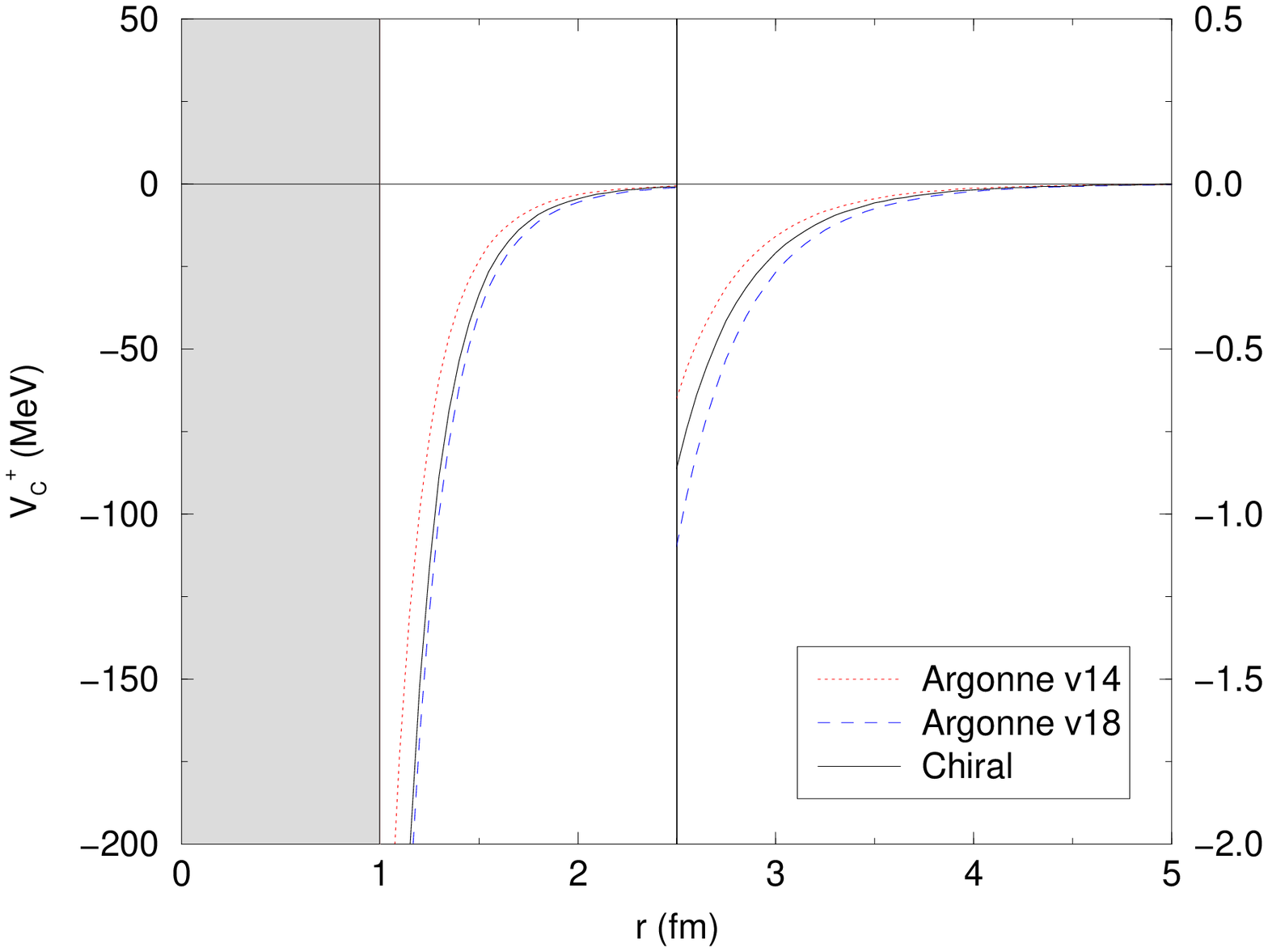}
\vspace{-2mm}
\caption{Isospin odd (left) and even (right) central components of the 
two-pion exchange potential.} 
\label{FB2}
\end{center}      
\end{figure}

The central components $V_C^-[\rar\cO(q^2)]$ and 
$V_C^+[\rar\cO(q^3)]$ are shown in fig.\ref{FB2}, 
and one has $|V_C^+| \sim 10 \, |V_C^-|$ in  regions of physical
interest, at odds with the predicted chiral hierarchy.
The numerical explanation for the magnitude of $V_C^+$ is that
it depends on large LECs generated by delta intermediate states.
Nevertheless, the prediction for $V_C^+$, which is by far the 
most important component of the nuclear force, is very good when compared 
with accurate phenomenological Argonne\cite{Arg} potentials.
Moreover, this agreement holds up to distances smaller than 1 fm, 
which correspond to momenta transferred $|t| > 20 \,M_\p^2$.
The empirical validity of results for $V_C^+$ is, thus, 
much wider than expectations allowed by chiral perturbation theory.

\section{NUCLEON SIGMA-TERM}

The structure of $V_C^+$ was discussed in refs.\cite{HRR,R01},
and about $70\%$ of its strength found to come from a term of the form
\bea
V_C^+(r) \sim -\, (2\,c_3/F_\p^2)\;\lb 2 - 4c_1/c_3 -  \bnb^2/M_\p^2 \rb \;
\tilde{\s}_{N}(r)\;,
\label{2.1}
\eea
where $\tilde{\s}_{N}$ is the long-distance part of the scalar
form factor in configuration space and 
the $c_i$ are usual LECs from the $\p N$ lagrangian.
An important role is played by $c_3$, which is large and dominated by 
$\D$ intermediate states.

The nucleon scalar form factor is defined in terms of the
symmetry breaking lagrangian as
\bea
\la N(p') | \sm \cL_{sb}\, | N(p) \ra =  \ub(p')\; u(p) \; \s_N(t) \;.
\label{2.2}
\eea
and has already been expanded to $\cO(q^4)$\cite{FM,BL}.
The leading $\cO(q^2)$ contribution 
comes from a tree diagram proportional to $c_1$, whereas
corrections at $\cO(q^3)$ and $\cO(q^4)$ are due to loops 
involving nucleon intermediate states and LECs.
The main features of these results were incorporated into a model
for the scalar form factor in configuration space\cite{sigma}, in which 
LECs at $\cO(q^4)$ are replaced by explicit $\D$ intermediate states.
The corresponding structure reads
\bea
\tilde{\s}_N(\br) &\!=\!& 
\lb - 4\, c_1\, \m^2\, \delta^3(\br) \; \rb_{\cO(q^2)} \;
+ \; \lb \;\tilde{\s}_{N}(r) \; \rb_{\cO(q^3)}^N \;
+ \; \lb \;\tilde{\s}_{N}(r) \;\rb_{\cO(q^4)}^\D \;,
\label{2.3}
\eea
the superscripts $N$ and $\D$ indicating intermediate propagators
in triangle diagrams.
These contributions are first evaluated in momentum 
space, by using $\cL_{sb} = F_\p^2 \, M_\p^2 \,\cos\theta$, 
where $\theta$ is the 
pion field, related to the usual unitary form by  
$U=\exp(i \btau \cdot \btheta) 
= \cos \theta + i \btau \cdot \hat{\btheta} \,\sin \theta$.
Results are then expressed as 
$ \s_N(t) = - F_\p^2 \,M_\p^2 \; \cos \theta(t) $.

Performing a Fourier transform and recalling that the vacuum expectation 
value of $\cL_{sb}$ is related to the light quark condensate by
$\la 0 |- \cL_{sb}\, | 0 \ra = \la 0 | \hat{m} (\ub u + \db d) | 0 \ra 
= -\,F_\p^2 \, M_\p^2$, the scalar form factor in coordinate space is
written as 
\bea
\check\s_N(r) = \la 0 | \hat{m} (\ub u + \db d) | 0 \ra \;
\cos \theta(r) \;.
\label{2.4}
\eea
The function $\cos\theta(r)$ describes the 
disturbance produced by the nucleon over the condensate and the
non-linear nature of pion interactions gives rise to the constraint
\bea
-1 \leq \check{\s}_N(r)/ \la 0 | \hat{m} (\ub u + \db d) | 0 \ra
\leq 1 \;.
\label{2.5}
\eea
Another condition over this ratio comes from the fact that the QCD 
ground state can take the form of either empty space or a 
quark-antiquark condensate.
In the present case, the boundary condition $\theta(r) \rar 0$ for 
$r \rar \infty$ ensures that the condensate remains undisturbed at 
large distances.
As one moves towards the nucleon, $\cos \theta$ {\em decreases},
indicating that it destroys the condensate.
This picture is compatible with the unitarity of the field $U$, which
correlates condensate and pion magnitudes and suggests that
the pion cloud of the nucleon 
is constructed at the expenses of the surrounding condensate, 
by means of a chiral rotation.
The model presented in ref.\cite{sigma} is based on the assumption 
that this process ends when all quark-antiquark pairs originally
present in the vacuum become excited, and a phase transition takes place
at the radius $R$ at which $\cos \theta(R) = 0$.
Formally, this corresponds to the condition
\bea
0 \leq \check{\s}_N(r)/ \la 0 | \hat{m} (\ub u + \db d) | 0 \ra
\leq 1 \;.
\label{2.6}
\eea

In configuration space, observables are calculated by integrating densities
over the entire volume.
In the case of the density $\check\s_N$, given by eq.(\ref{2.4}),
this would yield divergent results, since it does not vanish
in the limit $r\rar \infty$.
Therefore one shifts its origin, and works with a new function,
defined as
\bea
\tilde\s_N(r) \equiv \check\s_N(r) - \la 0 | \hat{m} (\ub u + \db d) | 0 \ra
= F_\p^2 \, M_\p^2 \; \lb 1-\cos\theta(r) \rb \;,
\label{2.7}
\eea
which describes  the nucleon cloud as a {\em deviation}
from the condensate. 
In practice, the function $\cos\theta$ cannot be calculated exactly
and one resorts to perturbation. 
This naturally yields a representation for $[1-\cos\theta]$
which vanishes at large distances and increases monotonically as one 
approaches the nucleon center.
At short distances, this representation becomes inadequate, 
since it is unbound and diverges at the origin.
In the model, this problematic region is excluded by the phase transition, 
for it assumes the perturbative representation for $\cos \theta$
in the range $R \leq r < \infty$ and $\cos \theta=0$ for $R < r$.

\begin{figure}[h]
\begin{center}
\hspace*{-4mm}
\includegraphics[width=0.35\columnwidth,angle=-90]{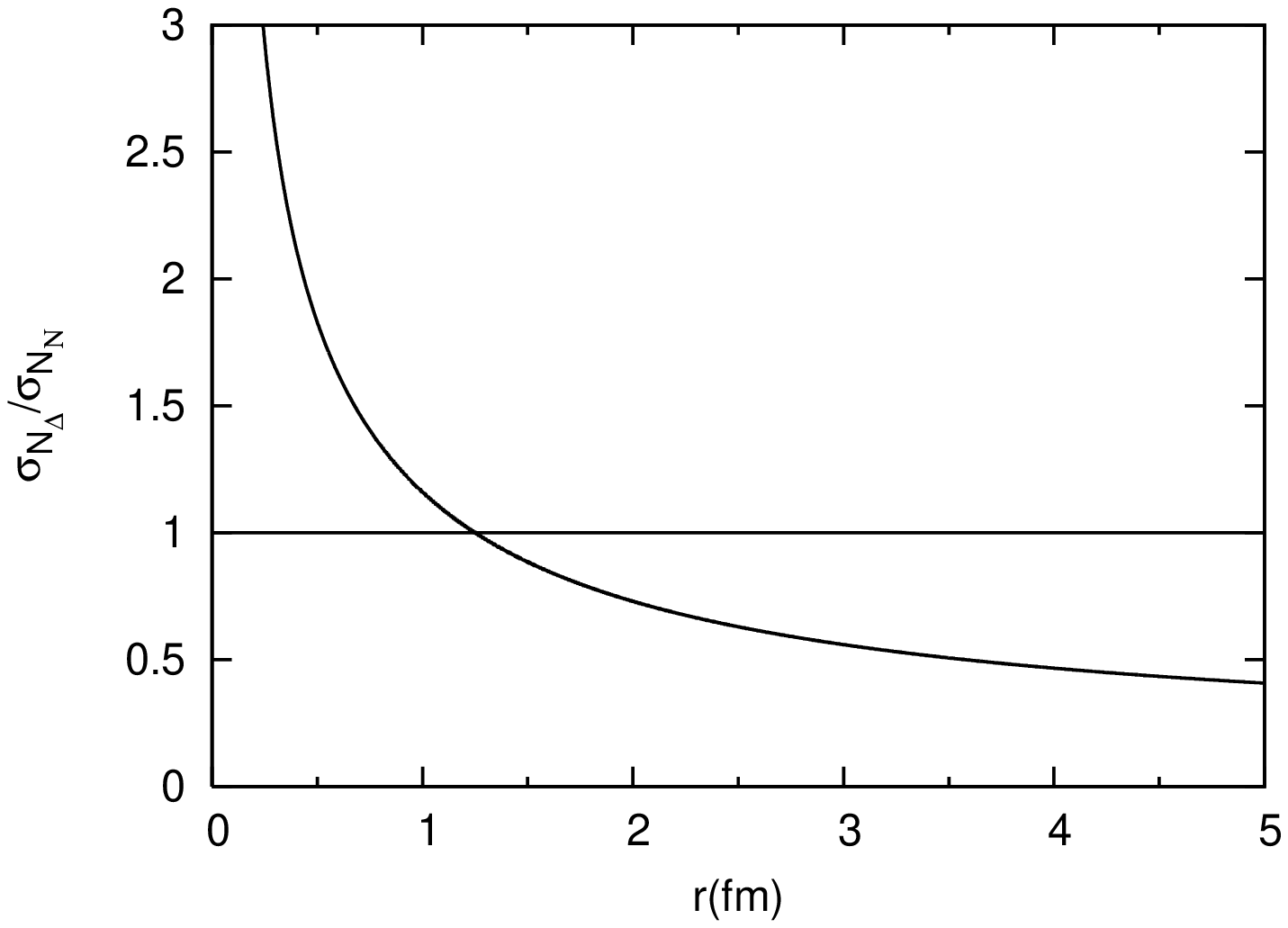}
\includegraphics[width=0.35\columnwidth,angle=-90]{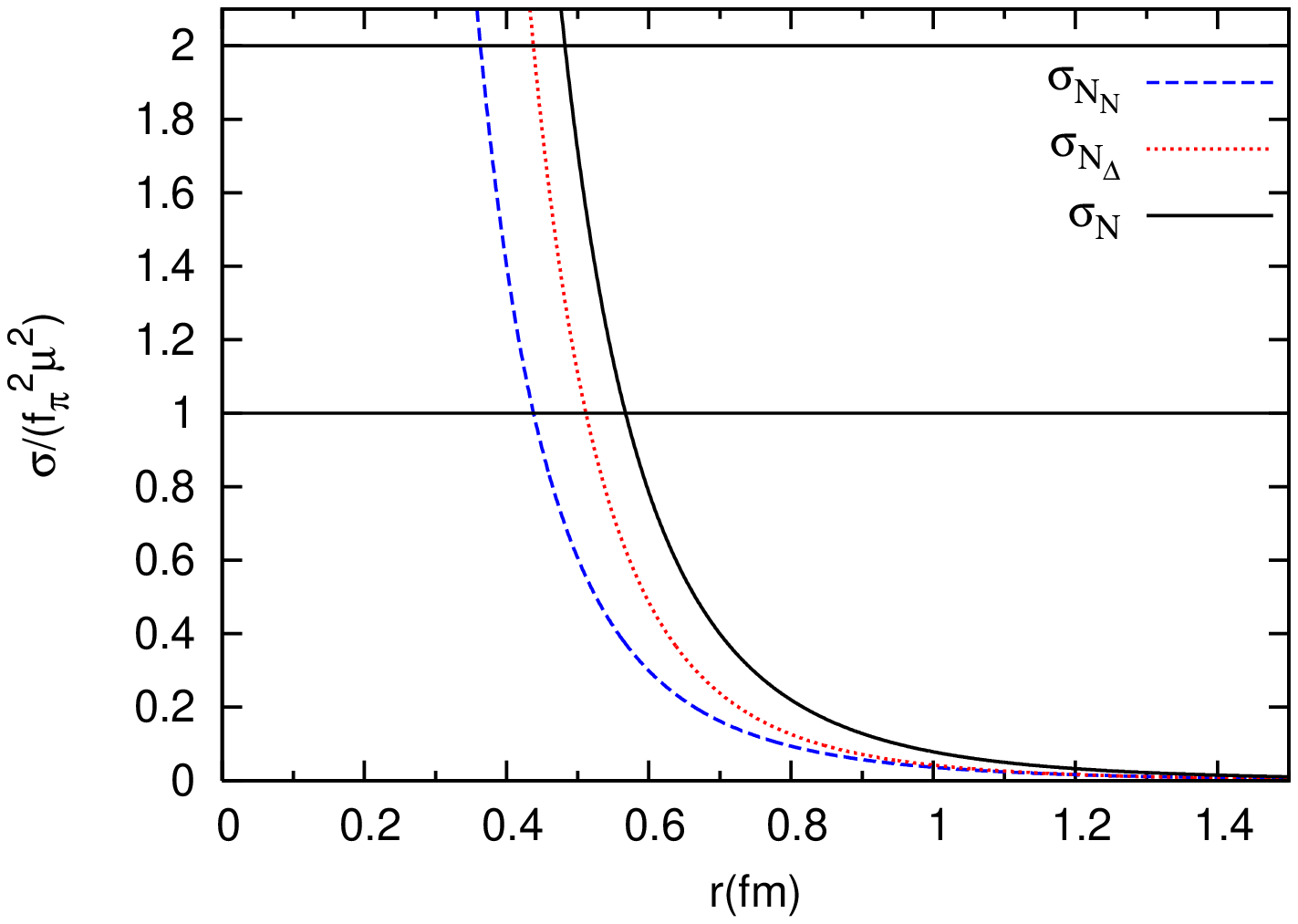}
\caption{Ratios $[\tilde{\s}_N(r)]^\D/[\tilde{\s}_N(r)]^N$ (left) 
and $\tilde{\s}_N(r)/(F_\p^2\,M_\p^2)=(1-\cos\theta)$ (right)
as functions of the distance to the nucleon center.}
\label{FB3}
\vspace{-3mm}
\end{center}      
\end{figure}

The roles played by $N [\rar \cO(q^3)]$ and $\D [\rar\cO(q^4)]$ 
intermediate states in eq.(\ref{2.3}) can be inferred from fig.\ref{FB3}.
The ratio $\lb \;\tilde{\s}_{N}(r) \; \rb^N /
\lb \;\tilde{\s}_{N}(r) \;\rb^\D$, given on the left,
shows that the hierarchy predicted by ChPT breaks down for distances 
smaller than 1.5 fm.
The right figure describes  the ratio 
$\tilde{\s}_N(r)/(F_\p^2\,M_\p^2)=(1-\cos\theta)$ inside this region,
together with individual $N$ and $\D$ contributions.
The phase transition is assumed to occur at the point 
$R\sim 0.6$ fm $(\rar |t|\sim 40\,M_\p^2)$,
where the black curve reaches the value 1.

In ref. \cite{sigma}, the nucleon $\s$-term has been evaluated 
by means of the expression
\bea
\s_N = \frac{4}{3} \p R^3 \;F_\p^2 M_\p^2
+ 4\p \int_{R}^\infty dr\;r^2\;\tilde{\s}_N(r) \;,
\label{2.8}
\eea
which yields 43 MeV$<\s_N<\;$49 MeV, depending on the value used
for the $\p N\Delta$ coupling constant, 
in agreement with the empirical value 45$\pm 8$ MeV \cite{GLS}.
Good results are also obtained for the $\Delta$ $\s$-term.

\section{SPACE-LIKE STRUCTURE OF THE PION}

Configuration space results for $V_C^+$ and $\s_N$ indicate that,
in these two instances, perturbative calculations are reliable at
high values of $|t|$.
This has motivated an exploratory study of pion properties
in a similar framework\cite{Rlec}, which yields good predictions for 
the mean square radius $(\la r^2 \ra_S^\p=0.59\,$fm$^2)$ and for one of
the LECs $(\bar{l}_4=4.3)$, with no free parameters.

The calculation departs from standard $\cO(q^4)$ results 
produced by Gasser and Leutwyler (GL) in 1984\cite{GL84},
and their notation is followed.
The pion scalar form factor is given by   
\bea
F_S(t) &\! = \!& \la \p(\bp') \,| \, \ub \,u + \db\, d \; |\,\p(\bp)\ra 
= 2 B \, \lc 1 + 2\,\m_\p + \frac{M_\p^2}{32\p^2\,F_\p^2} 
+ \frac{4 M_\p^2}{F_\p^2} \; l_3^r\right.
\nn\\[2mm]
&\! + \!& \left. \frac{(2\,t - M_\p^2)}{32 \p^2\, F_\p^2} \;L(M_\p,t) 
+ \frac{t}{F_\p^2}\,\lb l_4^r 
- \frac{1}{16\,\p^2} \,\lp \ln \frac{M_\p^2}{\m^2} + 1 \rp \rb \rc \;,
\label{3.1}
\eea
where $\m_\p = (M_\p^2/32 \p^2 F_\p^2) \ln (M_\p^2/\m^2)$,
the LECs are related to their scale-invariant 
counterparts by $l_3^r=-(\bar{l}_3+\ln M_\p^2/\m^2)/64\p^2$
and $l_4^r= (\bar{l}_4+\ln M_\p^2/\m^2)/16\p^2$,
$t=(p'-p)^2$, and  the loop function $L$ is related to the 
$\bar{J}$ in GL by $\bar{J}=L/(4\p)^2$.
In the Breit frame, the variable $t=-q^2$ is negative and one has 
\bea
&& L(M_\p,q) = 
 \s\; \ln \frac{\s - 1}{\s +1} + 2 \;,
\hspace*{5mm}
\s = \sqrt{1 + 4\,M_\p^2/q^2} \;.
\label{3.2}
\eea

The Fourier transform of $F_S(t)$ 
describes the spatial structure of the pion and reads 
\bea
\Ft_S(r) &\!= \!& 2 B \;
\lc \frac{\Lat(M_\p,r)}{32\p^2 F_\p^2} + \mathrm{ZRT}\rc \;,
\label{3.3}\\[2mm]
\Lat(M_\p,r) &\! = \!&
\int \frac{d^3q}{(2\p)^3}\; e^{-i\,\bq\cdot \br}\; 
[(2t-M_\p^2)\,L(M_\p,q)]
\label{3.4}\\[2mm]
 &\! = \!& \frac{M_\p^3}{\p\, r^2}\, \lb
\lp \frac{12}{M_\p^2 \, r^2} +7 \rp \,K_1(2M_\p r) 
+ \frac{12}{M_\p r}\;K_0(2M_\p r) \rb \;,
\label{3.5}
\eea
where the $K_i$ are Bessel functions and ZRT stands for zero range terms,
proportional to the $\d$-function and its derivatives.
The leading term in the quark condensate is given by 
$ \la 0 \, | \, \ub \,u + \db\, d \, |\,0\,\ra = - 2 B \,  F_\p^2 $,
and one writes 
\bea
\Ft_S(r) = - {\la 0 \, | \, \ub \,u + \db\, d \, |\,0 \ra} 
\; \frac{\Lat(M_\p,r)}{32\p^2 F_\p^4} + \mathrm{ZRT} \;.
\label{3.6}
\eea
At low-energies, the pion behaves as a Goldstone boson and,
far away from its center, the scalar form factor must be related 
to the surrounding quark-antiquark condensate by
$ \Ft_S(r) \rar N\; \la 0 \, | \, \ub \,u + \db\, d \, |\,0 \ra$,
where $N$ is a constant with dimension of mass.
However, eq.(\ref{3.6}) vanishes at large distances and,
as in the case of eq.(\ref{2.7}) for the nucleon, one has to perform 
a shift in the origin, defining a new function by
\bea
\check F_S(r) \equiv \tilde F_S(r) 
+ \la 0 | (\ub u + \db d) | 0 \ra
= N\; \la 0 \, | \, \ub \,u + \db\, d \, |\,0 \ra \; 
\lb 1 -\, \frac{\Lat(M_\p,r)}{N\,32\p^2 F_\p^4}  + \mathrm{ZRT} \rb \;.
\label{3.7}
\eea
This form is now suited for describing the behavior of the pion in the 
presence of the condensate. 
It is the analog to eq.(\ref{2.4}), with $\cos\theta$ replaced by 
$[1 -\, \Lat(M_\p,r)/(N\,32\p^2 F_\p^4)  + \mathrm{ZRT} ]$.
As in the nucleon case, it represents an undisturbed condensate at
large distances and decreases monotonically as one approaches the 
center of the pion.
For the same reasons as discussed in the previous section,
one assumes that this term is meaningful in the interval
\bea
0 \leq 
\lb 1 -\, \frac{\Lat(M_\p,r)}{N\,32\p^2 F_\p^4}  + \mathrm{ZRT} \rb
\leq 1 \;,
\label{3.8}
\eea
and that a phase transition occurs at a point $R$, such that
$1= \Lat(M_\p,R)/(N\,32\p^2 F_\p^4)$.
At smaller distances, the function $\Lat(M_\p,r)$ is replaced by the 
cut version $\T(r-R) \,\Lat(M_\p,r)$.
Zero range terms are then eliminated and one has 
\bea
\Fc_S(r) &\!=\!& -\,2\, B F_\p^2 \, 
\lc N - \T(r-R) \; \frac{\Lat(M_\p,r)}{32\p^2 F_\p^4} \rc \;.
\label{3.9}
\eea
This function does not vanish at infinity and, as in the nucleon case,
integration over entire space requires shifting the origin. 
The scalar form factor is then rewritten as
\bea
\Ft_S(r) = 2 \, B \,  
\lc \T(R-r)\; \frac{\Lat(M_\p,R)}{32 \p^2 F_\p^2} 
+ \T(r-R) \; \frac{\Lat(M_\p ,r)}{32 \p^2 F_\p^2} \rc \;,
\label{3.10}
\eea
after using the cutting condition to eliminate the factor $N$.

The scalar form factor in momentum space is given by the Fourier 
transform of this result and, for $t=0$, one has
$ F_S(0) = 4 \p \, \int dr \, r^2 \Ft_S(r)$.
At leading order, eq.(\ref{3.1}) yields $F_S(0)= 2\,B$ and 
one has the consistency condition
\bea
1 = 4 \p \, \int dr \, r^2 \; \frac{\Ft_S(r)}{2B} 
= \frac{M_\p^2}{16 \p^2 F_\p^2} \, 
\lb \lp \frac{20}{M_\p R} + \frac{14\, M_\p R}{3}\rp \,K_1(2M_\p R) 
+ 15 \, K_0(2M_\p R) \rb \;,
\label{3.11}
\eea 
which allows the cutting radius to be found.
The mean square radius (MSR), given by 
$ \la r^2 \ra_S^\p = 4\p \int dr\, r^4 \, [\Ft_S(r)/2B]$,
is an observable and reads
\bea
\la r^2 \ra_S^\p &\! = \!&  \frac{1}{80 \p^2 F_\p^2} 
\lb \lp 119 M_\p R + 14 M_\p^3 R^3 \rp K_1(2M_\p R) 
+ \lp 60 + 59 M_\p^2 R^2 \rp \,K_0(2M_\p R) \rb \;.
\label{3.12}
\eea

\section{RESULTS}

The cutting radius $R$ can be extracted from eq.(\ref{3.11}),
either numerically or by means of a perturbative expansion in $M_\p$.
As both results  coincide within $1\%$, one uses the latter.
Eq.(\ref{3.11}) becomes 
$ 1= 10 \, \lc 1
+ M_\p^2 R^2 \lb \lp \ln M_\p R + \g \rp/2
- 23/30 \rb \rc/(16 \p^2 F_\p^2 R^2) $
and yields    
\bea
R = \frac{\sqrt{10}}{4 \p F_\p}
\lc 1 + \frac{M_\p^2}{32 \p^2 F_\p^2} 
\lb 5 \lp \ln \frac{M_\p\sqrt{10}}{4\p F_\p} + \g \rp
- \frac{23}{3} \rb \rc \;,
\label{4.1}
\eea 
which corresponds to $R=0.500$ fm,
for $M_\p=139.57\,$MeV and $F_\p=92.4\,$MeV.
The MSR reads
\bea
\hspace*{-5mm}
\la r^2 \ra_S^\p &\!=\!& \frac{1}{16 \p^2 F_\p^2} 
\lc \frac{119}{10} - 12 \lp \ln \frac{M_\p\sqrt{10}}{4\p F_\p} + \g \rp 
- \frac{30 \,M_\p^2}{16 \p^2 F_\p^2}
\lb \lp \ln \frac{M_\p\sqrt{10}}{4\p F_\p} + \g \rp - \frac{61}{30} \rb \rc 
\label{4.2}
\eea
and produces $\la r^2 \ra_S^\p = 0.509$ fm$^2$, deviating
about $20\%$ from the precise estimate 
$\la r^2 \ra_S^\p = 0.61 \pm 0.04$ fm$^2$\cite{CGL}.

In ChPT, the MSR is related the LEC $\bar{l}_4$.
In the model, this LEC can be extracted by translating results 
back to momentum space.
The Fourier transform of the function $L(M_\p,q)$, 
given in eqs.(\ref{3.3} \ref{3.5}), involves ZRTs, which were
discarded in configuration space.
When one is interested in returning to momentum space, it is 
convenient to work with an extension of $L(M_\p,q)$, 
denoted by $L_e(q)$, and defined by the double integral
\bea
L_e(q) &\!=\!& \int_{M_\p^2}^{\m^2} db \,
\int_0^1 da\;\frac{1}{a\uma\,q^2 +b}
= L(M_\p,q) - \ln \frac{M_\p^2}{\m^2} - L(\m,q)\;,
\label{4.4}
\eea
where $\m$ is a scale. 
The function $L(\m,q)$ vanishes for large values of $\m$ and eq.(\ref{3.1})
becomes 
\bea
&& F_S(t) =  2B \, \lc \frac{(2\,t - M_\p^2)}{32 \p^2\, F_\p^2} \;L_e(t) 
+ \d \rc \;,
\label{5.3}\\[2mm]
&& \d = \lb 1
+ \frac{M_\p^2}{32 \p^2 F_\p^2} \lp 1 + \ln\frac{M_\p^2}{\m^2} \rp 
+ \frac{4 M_\p^2}{F_\p^2}\, l_3^r
+ \frac{t}{F_\p^2}\,\lp l_4^r 
- \frac{1}{16\,\p^2} \rp \rb \;.
\label{4.5}
\eea
Evaluating the Fourier transform and cutting the result 
at $r=R$, one has
\bea
\Ft_S(R,r) &\! = \!& 2B\, 
\lc \frac{\Lat_e(r)}{32 \p^2 F_\p^2} 
+ \,\T(R-r)\; \frac{\Lat_e(R) - \Lat_e(r)}{32 \p^2 F_\p^2} \rc \;,
\label{4.6}
\eea
where $\Lat_e(r) \equiv \Lat(M_\p,r)-\Lat(\m,r)$.
It is important to note that terms proportional to $\d$ gave rise to 
ZRTs and were discarded in the cutting procedure.

In returning to momentum space, $\Lat_e(r)$ becomes 
$(2\,t - M_\p^2) \, L_e(t)$ again, and a {\em new} factor $\d$ is created 
by the Fourier transform of the term proportional to $\T(R-r)$
in eq.(\ref{4.6}). 
The functions $\Lat_e(r)$ are expressed in terms of Bessel functions
$K_i$ and an explicit calculation produces
\bea
\hspace*{-9mm}
&& 4\p\, \int_0^R dr \; e^{i\,\bq\cdot \br}\, r^2
\; \frac{\Lat_e(R) - \Lat_e(r)}{32 \p^2 F_\p^2} 
= 
\lb 1 - \frac{M_\p^2}{32\,\p^2 F_\p^2}\,\ln\frac{M_\p^2}{\m^2}
+ \frac{t}{16 \, \p^2 F_\p^2}\,
\lp \frac{31}{15} - 2 \, \lp \ln \m R + \g \rp \rp \rb , 
\label{4.7}
\eea
for low values of $q^2$.
Comparing eqs.(\ref{4.5}) and (\ref{4.7}), one finds
\bea
&& 4 \, l_3^r = - \frac{1}{16\p^2} \,
\lb \frac{1}{2} + \ln \frac{M_\p^2}{\m^2} \rb 
\rar \; \bar{l}_3 = 1/2 \;,
\label{4.8}\\[2mm]
&& l_4^r = \frac{1}{16 \p^2} \lb \frac{46}{15} 
-2 (\ln M_\p R +\g) + \ln \frac{M_\p^2}{\m^2} \rb
\rar \; \bar{l}_4 = \frac{46}{15} - 2 \lp \ln M_\p R  +\g \rp \;.
\label{4.9}
\eea
Both results contain the correct $\ln M_\p^2/\m^2$ structure, 
but that concerning $\bar{l}_3$ cannot be trusted, since it is 
based on the approximation $F_S(0) = 2\,B$ used in eq.(\ref{3.11}).   
The prediction for $\bar{l}_4$ is consistent with the 
$\la r^2 \ra_S^\p$ given in eq.(\ref{3.12}), since it follows the
relation\cite{GL84}
$\la r^2 \ra_S^\p =  3 \lb \bar{l}_4 - 13/12 \rb/(8 \p^2 F^2) $
at leading order.
Chiral perturbation theory at two loops predicts\cite{CGL} 
$\bar{l}_4 = 4.4 \pm 0.2$.
Using the value $R=0.500\,$fm produced by eq.(\ref{4.2}), one finds
$\bar{l}_4 = 3.99$.

\vs{2mm}

Results for $\la r^2 \ra_S^\p$ and $\bar{l}_4$ are improved by
the inclusion of scalar mesons.
One follows the work of Ecker, Gasser, Pich and De Rafael \cite{EGPR}
and adopts their values 
$M_S=M_{S_1}\equiv m=980\,$MeV,
$c_d = \sqrt{3}\; \tilde{c}_d= 32\,$MeV, and
$c_m = \sqrt{3}\; \tilde{c}_m= 42\,$MeV.
In the expressions that follow, terms already displayed 
previously are denoted by $[\cdots]$. 
The scalar form factor in momentum space now reads
\bea
F_S(t)  
&\!=\!& 2 B\, \lc [\cdots] 
- \frac{4\,c_m}{F_\p^2} \lb c_d + 
\frac{c_d\,m^2 + 2(c_m - c_d)M_\p^2}{t-m^2}\rb \rc 
\label{4.10}
\eea
and corresponds to 
\bea
\Ft_S(r) =  2 B \;
\lc [\cdots] + \frac{E\,e^{-mr}}{4\p \,r} + \mathrm{ZRT}\rc \;,
\hspace*{10mm}
E = \frac{4 c_m}{F_\p^2} \lb c_d\,m^2 + 2(c_m - c_d)M_\p^2 \rb \;.
\label{4.11}
\eea
Cutting the integrand at the radius $R$, one finds the new version 
of eq.(\ref{3.10}), given by
\bea
\Ft_S(r)=   
\lc \T(R-r)\; \lb \frac{\Lat(M_\p,R)}{32 \p^2 F_\p^2}
+ \frac{E\,e^{-mR}}{4\p \,R} \rb 
+ \T(r-R) \; \lb \frac{\Lat(M_\p,r)}{32 \p^2 F_\p^2} 
+ \frac{E\,e^{-mr}}{4\p \,r} \rb \rc \;.
\label{4.12}
\eea
Imposing the spatial integral of $\Ft_S(r)$ to be equal to $F_S(0)$,
the condition for determining the cutting radius becomes
\bea
1 &\! = \!& \frac{M_\p^2}{16 \p^2 F_\p^2} \, 
\lb \lp \frac{20}{M_\p R} + \frac{14\, M_\p R}{3}\rp \,K_1(2M_\p R) 
+ 15 \, K_0(2M_\p R) \rb 
\nn\\[2mm]
&\! + \!& \frac{E}{m^2} \lb 1 + mR + \frac{m^2 R^2}{3} \rb \;e^{-mR} \;,
\label{4.13}
\eea 
and yields $R=0.570\,$fm $(\rar |t|\sim 40\,M_\p^2)$.
The mean square radius is now given by
\bea
\la r^2 \ra_S^\p &\! = \!& [\cdots]
+ \frac{E}{m^4} 
\lb 6 + 6 mR +3 m^2 R^2 +m^3 R^3 + \frac{m^4 R^4}{5}\rb \; e^{-mR}
\label{4.14}
\eea
and has the value $\la r^2 \ra_S^\p = 0.591$ fm$^2$,
to be compared with\cite{CGL}  $\la r^2 \ra_S^\p = 0.61 \pm 0.04$ fm$^2$.
The procedure for obtaining $\bar{l}_4$ is the same as before, 
and one evaluates the integral 
\bea
&& 4\p\,\int dr \; e^{i\,\bq\cdot \br}\;r^2 \;
\lb \frac{\Lat_e(R) - \Lat_e(r)}{32 \p^2 F_\p^2} 
+ \frac{E}{4\p} \lp \frac{e^{-mR}}{R} - \frac{e^{-mr}}{r} \rp \rb
\nn\\[2mm]
&& = 
\lb 1 + [\cdots]
+ t \, \frac{E}{m^4} \lp 1 + mR + \frac{m^2 R^2}{2} 
+ \frac{m^3 R^3}{6} + \frac{m^4 R^4}{30} \rp \;
e^{-mR} \rb \;,
\label{4.15}
\eea
which yields 
\bea
\bar{l}_4 &\! = \!& \frac{46}{15} - 2 \lp \ln M_\p R  +\g \rp 
+ \frac{64 \p^2 \, c_m \lb c_d + 2(c_m - c_d) M_\p^2/m^2 \rb }{m^2}\,
\nn\\[2mm]
&\! \times \!& 
\lp 1 + mR + m^2 R^2/2 + m^3 R^3/6 + m^4 R^4/30 \rp \;e^{-mR} \;.
\label{4.16}
\eea
Numerically, this corresponds to $\bar{l}_4 = 4.26$, 
within the error bars of the precise result $\bar{l}_4 = 4.4 \pm 0.2$ 
derived by Colangelo, Gasser and Leutwyler\cite{CGL}.
In the alternative notation $\bar{l}_4 = \ln \Lambda_4^2/M_\p^2$,
one finds $\Lambda_4=1.178\,$GeV.



\begin{thebibliography}{99}

\bibitem{Epelbaum} for a comprehensive review, see E. Epelbaum,
H-W. Hammer and U-G. Meissner, preprint nucl-th/0811.1338.

\bibitem{W-ChPT} S. Weinberg, Physica A {\bf 96}, 327 (1979),
Phys. Lett. B {\bf 251}, 288 (1990), 
Nucl. Phys. B {\bf 363}, 3 (1991).

\bibitem{R-Osaka} M.R. Robilotta, Mod. Phys. Lett. A {\bf 23}, 2273 (2008).

\bibitem{CDR} H. T. Coelho, T. K. Das, and M. R. Robilotta, 
Phys. Rev. C {\bf 28}, 1812 (1983).

\bibitem{HR} R. Higa and M.R. Robilotta, Phys. Rev. C {\bf 68}, 024004 (2003).

\bibitem{HRR} R. Higa, M.R. Robilotta and C. A. da Rocha, 
Phys. Rev. C {\bf 69}, 034009 (2004).

\bibitem{RR94} C. A. da Rocha and M. R. Robilotta, Phys. Rev. C {\bf 49}, 
1818 (1994).

\bibitem{BL} T. Becher and H. Leutwyler, Eur. Phys. Journal C {\bf 9}, 643 
(1999);
JHEP {\bf 106}, 17 (2001).

\bibitem{Arg} R.B. Wiringa, R.A. Smith,and T.L. Ainsworth,
Phys. Rev. C {\bf 29}, 1207 (1984);
R.B. Wiringa, V.G.J. Stocks, and R. Schiavilla, 
Phys. Rev. C {\bf 51}, 38 (1995).

\bibitem{R01} M. R. Robilotta, Phys. Rev. C {\bf 63}, 044004 (2001).

\bibitem{FM} N. Fettes and U-G. Meissner, Nucl. Phys. A {\bf 693}, 
693 (2001);
{\em ibid.} A {\bf 676}, 311 (2000).

\bibitem{sigma} I.P. Cavalcante, M.R. Robilotta, J. S\'a Borges, 
D.O. Santos, and G.R.S. Zarnauskas, Phys. Rev. C {\bf 72}, 065207 (2005).

\bibitem{GLS} J. Gasser, H. Leutwyler and M.E. Sainio, 
Phys. Lett. B {\bf 253}, 252 (1991); 
{\bf 253}, 260 (1991).

\bibitem{Rlec} M. R. Robilotta, submitted for publication.

\bibitem{GL84} J. Gasser and H. Leutwyler, Ann. Phys. {\bf 158}, 142 (1984).

\bibitem{CGL} G. Colangelo, J. Gasser and H. Leutwyler, 
Nucl. Phys. B {\bf 603}, 125 (2001)
 
\bibitem{EGPR} G. Ecker, J. Gasser, A. Pich and E. De Rafael, 
Nucl. Phys. B {\bf 321}, 311 (1989). 






\end{thebibliography}
\end{document}